\documentclass[12pt,preprint]{aastex}

\def\k{{mnl}}
\def\kk{{m, n, l}}
\def\kms{km$\,$s$^{-1}$}
\def\LH{LH}
\def\ml{M$_\odot/$L$_{V,\odot}\;$}
\def\rhods{\hbox{$\times10^7$M$_\odot/$kpc$^3$}}
\def\gml{${\bar\Gamma_V}$}
\newcommand{\ve}[1]{\mathbf{#1}}

\shorttitle{Mass distribution of dSph galaxies}
\shortauthors{Wu}

\begin{document}
\title{The mass distribution of dwarf spheroidal galaxies from stellar kinematics: Draco, Ursa Minor and Fornax}
\author{Xiaoan Wu} 
\affil{Princeton University Observatory, Peyton Hall}
\affil{Princeton, NJ 08544-1001, USA}
\email{xawn@astro.princeton.edu}

\begin{abstract}
We model three dSph galaxies, Draco, Ursa Minor and Fornax, as 
axisymmetric stellar systems embedded
in spherical dark-matter potentials, which are in
dynamical equilibrium without significant external tidal forces.
We construct non-parametric two- and
three-integral models that match the observed surface-density profiles and
the current kinematic samples of $\sim150-200$ stars per galaxy; 
these models naturally produce the so-called
``extra-tidal extensions", which had previously been suggested as
evidence of tidal stripping.
Isochrone, NFW and power-law models fit the data, but
we strongly rule out any centrally condensed mass distribution 
like a dominant central black hole, as well
as constant-density and (for Draco and Fornax) constant mass-to-light ratio models.
The average V-band mass-to-light ratio is $400\pm80$ \ml 
within $0.75$ kpc for Draco,
$580^{+140}_{-100}$ \ml within $1.1$ kpc for Ursa Minor and
$25^{+7}_{-5}$ \ml within $2.5$ kpc for Fornax.
Two-integral models fit the data almost as well as three-integral
models; in contrast to previous suggestions we do not find that
anisotropy contributes substantially to the
high mass-to-light
ratios in these dSph galaxies.

\end{abstract}


\section{Introduction}
\label{sec:intro}

Dwarf galaxies include dwarf irregular (dIrr) and dwarf spheroidal (dSph) galaxies. Their
luminosities \citep[$10^3 -10^8 L_\odot$, see e.g.][]{Bel07} are 
much smaller than the characteristic luminosity 
$L^*\sim2\times10^{10} L_\odot$ for giant galaxies. The faintest dSph galaxies have luminosities
comparable to globular clusters, but are more extended (half-light radius $\sim 0.5$ kpc compared to 
$\sim 0.005 $ kpc for globular clusters) and thus have much lower surface brightness.

The formation history of dwarf galaxies is not well-understood. 
The standard theory is that dwarf galaxies formed at the centers of subhalos orbiting
within the halos of giant galaxies. However, the
standard $\Lambda$CDM model predicts a much larger number
of subhalos than the observed number of dwarf galaxies. There are $\sim 50$ known 
dwarf galaxies in the Local Group, compared to 300 predicted subhalos with circular velocity $\sim 10-20$ \kms \citep{Kly99}.

\citet{Aar83} was the first to determine the velocity dispersion of a dSph galaxy, using three
stars in the Draco dwarf (6.5 \kms). Aaronson's result 
was later confirmed by more extensive observations
\citep{Ols96, Wil04}. It is found that
all dwarf galaxies have 
central velocity dispersions $\sim 6-25$ \kms \citep{Mat98}. These dispersions are believed to reflect motions
of the stars in the gravitational potential of the galaxy rather than (say) atmospheric motions or
orbital motion due to a binary companion
\citep{Ols96,Mat98}. 
If the systems are in dynamical equilibrium, the mass derived from these velocity dispersions
is much larger than the
stellar mass, resulting in a V-band mass-to-light ratio ranging from 7 to 500 \ml
\citep{IH95, Wil02, Pal03, Kle05, Lok05}. 
Therefore, dSph galaxies are probably the darkest objects ever observed in the universe, and
thus provide unique probes of the dark-matter distribution on small scales. 

Two possible alternatives to dark matter in dSph galaxies have been proposed. The first 
is Modified Newtonian Dynamics \citep[MOND,][]{Mil83}. However,
MOND has difficulty reproducing the different mass-to-light ratios of different
dSph galaxies, mostly because MOND is based on a characteristic acceleration whereas the dSph galaxies seem
to have a characteristic velocity dispersion instead. Moreover, MOND is difficult to reconcile with the
power spectrum of fluctuations in the CMB \citep{Spe06}. 
The second alternative is tidal forces from the host galaxies.
Tides have changed the structure and internal kinematics of the
 Sagittarius dwarf galaxy \citep{Iba94, Iba01}.
So it is natural to postulate that other dSph galaxies are also undergoing tidal disruption, which leads
to the high velocity dispersions \citep{Kro97, Mar01, Go03}. However, (i) Sagittarius is much
closer to the center of the Milky Way than other dwarf galaxies; (ii) 
tides should introduce strong velocity shear and dSph galaxies rotate slowly if at all; (iii)
no mechanism of increasing 
velocity dispersions by tides appears to work for all dwarf galaxies \citep{Mat98, Kle03}. 
There are a few pieces of putative evidence that tides may influence the structure of
Ursa Minor and Draco \citep{Mar01, Pal03, Mun05}. 
For example, ``extra-tidal extensions'' \citep[e.g.,][]{IH95, Wil04} are found beyond a ``tidal radius'' by
fitting a King profile \citep{King66} to stellar number density profiles. However,
the King profile was intended to fit relaxed,  constant mass-to-light ratio systems such as
globular clusters \citep{Tra95}, and
perhaps the profile can not be applied to dSph galaxies.

In this paper we model the mass distribution of three dSph galaxies, 
specifically, Draco, Ursa Minor 
and Fornax. Since we focus on the inner parts ($\leq 2$ kpc) which
are very regular \citep{IH95} and not significantly affected by tidal stripping \citep{Rea06},
we assume that the inner part is in dynamical
equilibrium with no significant external tidal forces, and test whether this assumption is self-consistent with a given mass distribution. 

A wide variety of approaches have been used to model the kinematics of dSph galaxies.
Different analysis techniques and assumptions have
led to different results. Taking Draco as an example (see also Figure \ref{fig:draco}), if we calibrate the results
assuming a V-band luminosity of $1.8\times 10^5 L_\odot$ \citep{IH95}, 
\citet{Ode01} derive a V-band mass-to-light ratio
of \gml$=195\pm56$ \ml within 1.2 kpc
using a King model.  
\citet{Wil02} and \citet{Kle02} assume parametric forms for both potential and distribution (hereafter DF), and
model Draco as a near-isothermal sphere ($v_{circ}\propto r^{0.17}$), with \gml$= 440\pm240$ \ml 
within 0.7 kpc, more than a factor of two larger
than derived by \citet{Ode01} at a larger radius.
By fitting velocity moments, \citet{Lok02} investigates the dark matter distribution with 
parametric DFs and the assumption that dSph galaxies are spherical and derives
\gml$=175\pm35$ \ml within 0.4 kpc.
\citet{Mas04} propose that the mass of Draco could range from $7\times10^7$ to $5\times10^9 M_\odot$, assuming an
NFW \citep{Nav97} potential, which has \gml$=540\pm170$ and $510\pm160$ at $r=0.75$ and 1.1 kpc, respectively.
\citet{Lok05} parametrize anisotropy by a constant anisotropy parameter $\beta$ to
fit velocity moments and found 
the mass-to-light ratio was $\sim370$ \ml and almost constant with radius for Draco. 
\citet{Wan05} have determined dark matter distributions, using a sophisticated non-parametric mass distribution,
assuming spherical symmetry. However, they assume that the distribution function is
isotropic in velocity space, an assumption that can introduce much larger errors than the ones that
are controlled by the use of non-parametric mass distribution. We believe that
a better method is to use a non-parametric distribution function and a parametric potential. 

The apparent ellipticities\footnote{If the apparent semi-major and minor axes of isopleths are
$a$ and $b$, then the apparent ellipticity $e_a=1-b/a$. Intrinsic ellipticity $e_i$ is defined similarly, i.e.,
it is equal to $e_a$ while the galaxy is seen edge-on.} 
of the stellar distributions of
dSph galaxies vary from 0.13 (Leo II) to 0.56 (Ursa Minor) \citep{Mat98}. 
However, the gravitational potential is dominated
 by the dark halo in these galaxies, and there is no reason to suppose that the dark halo shape is the same as
the shape of the stellar distribution. Hence for simplicity, we assume that the galactic potential is spherical.
In addition, we assume that the distribution of stars in the dSph galaxy is 
axisymmetric.
Other methods designed to model axisymmetric galaxies \citep[e.g.,][]{Cre99} are more general
in that they allow both the potential and the distribution function to be axisymmetric; however,
these are designed to work for a continuous, rather than discrete, observed distribution of stars.
We describe and test our method in \S\ref{sec:method} and \S\ref{sec:sim}, 
respectively.
Then we derive the mass density profiles for Draco, Ursa Minor and Fornax in \S\ref{sec:res}.
We discuss the results in \S\ref{sec:dis} and conclude in \S\ref{sec:con}.

\section{Method}
\label{sec:method}

We introduce a Cartesian coordinate system, with origin at the center of the dwarf galaxy and 
$z-$axis along the line of sight. 
The $x$ and $y$-axes are the apparent major and minor axes in the plane of the sky, 
respectively, which can be determined from
isodensity contours of the surface number density profile (isopleths).

 Then the
distribution function may be written as $f(E, L, L_{z'})$, where $E$ is energy, $L$ is scalar
angular momentum, the $z'-$axis is
the intrinsic symmetry axis of the stellar distribution and $L_{z'}$ is the angular momentum along the
$z'$-axis. Therefore, the $z'$-axis lies in the $y-z$ plane. For an observational data set 
of positions and line-of-sight velocities $\{x_i, y_i, v_{zi}\}$, 
we model $f(E, L, L_{z'})$ and determine both the parameters of the potential and the inclination
$\theta$, which is the angle between the $z'$-axis and 
the $z$-axis.

\citet[hereafter Paper I]{Wu06} have used maximum likelihood analysis to deal with the analogous problem
for spherically symmetric DFs in a spherical potential, where the DF is a function
of two integrals, $E$ and $L$. 
It is straightforward to generalize this method to the case of an axisymmetric DF $f(E, L, L_{z'})$. 
We define a probability function,
\begin{eqnarray}
g(x, y, v_z)=\int f(E, L, L_{z'}) dv_x dv_y dz
\end{eqnarray}
which gives the probability density to observe a star at $(x, y)$ with a line-of-sight velocity $v_z$.
Note that $g$ is also a function of  $\theta$ and the assumed gravitational potential $\Phi$.
We maximize
the log likelihood function
\begin{eqnarray}
LH(\Phi, f, \theta)=\sum _i \ln g(\{x_i, y_i, v_{zi}\}|\Phi, f, \theta), \label{eq:LHg}
\end{eqnarray} 
assuming parametrized potentials and non-parametric DFs as follows.

For the potential, we assume an analytical form $\Phi(r, \ve{X})$
and infer the parameters $\ve{X}$. We investigate six possible forms.
\begin{enumerate}

\item
 An isochrone model, 
\begin{eqnarray}
\Phi(r)=-\frac{GM}{b+\sqrt{b^2+r^2}}, \;\; \ve{X}=\{M, b\},\label{eq:iso}
\end{eqnarray}
where $M$ is the total mass. 

\item
The \cite{Nav97} density-potential pair,
\begin{eqnarray}
\rho(r)&=&\frac{\rho_n}{\displaystyle{\frac{r}{r_n}}\left (1+\displaystyle{\frac{r}{r_n}} \right)^2}, \nonumber\\
\Phi(r)&=&-4\pi G\rho_nr_n^2\frac{\ln \left (1+\displaystyle{\frac{r}{r_n}} \right)}{\displaystyle{\frac{r}{r_n}}}, \label{eq:NFW}
\end{eqnarray}
where $r_n$ is the concentration radius and $\ve{X}={\{\rho_n, r_n\}}$.

\item
 A power-law  density-potential pair
\begin{eqnarray}
\rho(r)&=&\rho_0\left (\frac {r}{r_0}\right )^{-\alpha}, \nonumber\\
\Phi(r)&=&\frac{4\pi G\rho_0 r_0^2}{(2-\alpha)(3-\alpha)}\left (\frac {r}{r_0} \right )^{2-\alpha} \;\;\;(\alpha<3),  \label{eq:powerlaw}
\end{eqnarray}
where $r_0$ is an arbitrary radius,
and then $\ve{X}={\{\rho_0, \alpha\}}$. 

\item
A constant-density model, which is a special case of the power-law model
with $\alpha=0$ and $\ve{X}={\{\rho_0\}}$.

\item
 A constant mass-to-light ratio model (see \S\ref{sec:ml}). 

\item
A model in which the potential is dominated by a central black hole of mass $M_{bh}$,
\begin{eqnarray}
\Phi(r)&=& -GM_{bh}/r, \;\; \ve{X}=\{M_{bh}\}. \label{eq:bh} 
\end{eqnarray}

\end{enumerate}

For comparison with earlier analysis by other authors, we shall also fit the data
  using a more conventional method: fitting to the \citet{King66}
  three-parameter family of isotropic stellar systems, by minimizing
  $\chi^2$ for the surface-density and velocity-dispersion profiles
  (see \S\ref{sec:king}).

There is little rotation in dSph galaxies. Therefore, we do not attempt to model the observed rotation: 
we simply assume $f$ is an even function of $L_{z'}$.
Furthermore, the non-rotating model has even symmetry around the $x$ and $y$-axes, i.e., 
$g(x, y, v_z)=g(|x|, |y|, |v_z|)$, so
we can assume that $x, y, v_z$ are all positive and $0\leq\theta\leq\pi/2$. 
These simplifications increase the computational efficiency significantly.

To construct the DF, we divide the $\{E,L, |L_{z'}/L|\}$ space into $N_E\times N_L \times N_{z'}$ bins, which
are denoted by the triple index $\k$, $m=1, ..., N_E$,
$n=1, ..., N_L$, $l=1, ..., N_{z'}$. The bins are mutually exclusive and 
cover all of the allowed $E-L-L_{z'}/L$ space, as well as some unallowed regions.
In our calculation, bin $\k$ is defined by
\begin{eqnarray}
E_{m-1}&\leq E <& E_{m}, \label{eq:Ebin}\\
\frac{n-1}{N_L}L_c^2(E_{m}) &\leq L^2 <&\frac{n}{N_L}L_c^2(E_{m}), \label{eq:Lbin} \\
\cos \gamma_{l-1} \geq  \left | \frac{L_{z'}}{L} \right |> \cos \gamma_l, &\;\hbox{where}\; &
\gamma_l  \equiv  \frac{ul\pi}{2N_{z'}}. \label{eq:Lzbin}
\end{eqnarray}
Here $E_{m-1}$ and $E_{m}$ are the lower and upper limits of the energy in the bin, and $L_c(E)$ is the
maximum angular momentum at energy $E$, corresponding to a circular orbit. To achieve 
uniform spatial resolution for different potential models, we choose the boundaries of energy bins $\{E_{m}\}$ as 
the potentials at a set of evenly separated radii. The limits in $L_{z'}/L$ are
chosen so that the resolution in $\gamma$, the angle between the $z'-$axis and the vector normal to the orbital plane 
(the orbital inclination) is uniform. Note that for spherical systems, the natural choice is
uniform in $\cos \gamma$, not $\gamma$. However, when modeling flattened systems, this choice yields too few bins near $\gamma=0$. 
The factor $u$ determines the maximum value for $\gamma$ and must
be 1 for the bins to cover all allowed regions. However, it
can be chosen to be $<1$ to
model a galaxy with very flat (disk-like) stellar part with higher computational efficiency.

Assuming that the DF is constant in bin $\k$ and zero elsewhere,
we can calculate a probability distribution in $(x, y, v_z)$ space (cf. Paper I), 
\begin{eqnarray}
g_\k(x, y, v_z|\ve{X}, \theta)\propto \int _{\hbox{bin}\;\k} dv_x dv_y dz. \label{eq:gmnl}
\end{eqnarray}
we normalize
each $g_\k(x, y, v_z|\ve{X}, \theta)$ so that $\int g_\k(x, y, v_z|\ve{X}, \theta) dx dy dv_z=1$.

Assuming the non-negative weights in the bins are $\ve{W}=\{w_\k\}$, we have
\begin{eqnarray}
\sum _\kk w_\k &=& 1, \nonumber\\
g(x, y, v_{z})&=&\sum _\kk w_\k g_\k(x, y, v_{z}), \nonumber\\
LH(\ve{X}, \ve{W}, \theta) &=&\sum_i \ln \left ( \sum _\kk w_\k g_\k(\{x_i, y_i, v_{zi}\}|\ve{X}, \theta) \right ), \label{eq:LH}
\end{eqnarray}
where $LH$ is the log of the likelihood. 
Notice that $g(x, y, v_{z})$ has been normalized since all $g_\k(x, y, v_{z})$ are normalized.
For a fixed inclination $\theta$ and fixed potential specified by the parameters $\ve{X}$, we
maximize $LH(\ve{X}, \ve{W}, \theta)$ with respect to $\ve{W}$,
then vary $\ve{X}$ and $\theta$ to search for a global maximum $LH(\ve{X}, \ve{W}, \theta)$. More generally,
the posterior probability of $(\ve{X}, \theta)$ is given by
\begin{eqnarray}
P(\ve{X}, \theta)\propto P_\ve{X}P_{\theta} \max_\ve{W}\left ( P_\ve{W} e^{LH(\ve{X}, \ve{W})} \right ), \label{eq:xv}
\end{eqnarray}
where $P_{\theta}$, $P_\ve{X}$ and $P_\ve{W}$ are prior probabilities. We use
\begin{eqnarray}
P_{\theta}=\sin\theta, \label{eq:ptheta}
\end{eqnarray}
corresponding to an isotropic distribution of orientations. 
Our rule for choosing $P_X$ is to 
assign a uniform probability distribution for dimensionless parameters and a uniform probability
distribution in log scale for dimensional parameters. We choose $P_W=1$ 
in this paper; that is, in contrast to Paper I we do not use regularization to smooth the DF (in Paper I, we found that  the errors in derived potential parameters for
samples of a few hundred stars were dominated by statistical noise
rather than different regularization levels).

It is straightforward to account for observational errors in velocities and observational cuts on the
sample volume in velocity and position on the sky (velocity cuts are normally applied to eliminate interlopers). The former
is dealt with by convolving $g_\k(x, y, v_z)$ with a Gaussian error profile. The latter requires us to 
normalize $g_\k(x, y, v_z)$ within the cuts in velocity and position.

If the surface density of the kinematic sample is proportional to the overall surface density of stars, we use Equation (\ref{eq:LH}). 
If not, the distribution
of positions contains no information, so in the likelihood function
we use the conditional probability
\begin{eqnarray}
h(v_{zi}|x_i, y_i, \ve{X}, \theta)=\frac{g(x_i, y_i, v_{zi}|\ve{X}, \theta)} 
                                          {\int g(x_i, y_i, v_{z}|\ve{X}, \theta) dv_z}. 
\end{eqnarray}
In this paper, we use Equation (\ref{eq:LH}) for the calculations in this paper.

An observed surface number density profile may be available for many
more stars, over a larger survey area than the kinematic sample. We can
easily add this as an additional constraint.
Assume the apparent ellipticity 
of the stellar part is $e_a$, and that its surface number density profile along
the $x-$axis is
$\Sigma_{obs}(R_{e})$ with error $\sigma_\Sigma(R_{e})$, where $R_{e}$ is the distance along the $x-$axis.
We maximize the quantity
\begin{eqnarray}
Q &\equiv&\LH-\frac{1}{2}\chi ^2, \label{eq:Qwp}\\
\chi ^2&=&\sum_j \frac {(\Sigma(R_{ej})-\Sigma_{obs}(R_{ej}))^2}{\sigma_\Sigma(R_{ej})^2}, \label{eq:x2}\\
R(\phi, R_e)&=&\frac{R_e}{\sqrt{\cos^2 \phi+\sin^2\phi/(1-e_a)^2}}, \nonumber\\
\Sigma(R_e) &\propto &\int \frac{g(R\cos\phi, R\sin\phi, v_z)}{\cos^2 \phi+\sin^2\phi/(1-e_a)^2} dv_z d\phi \nonumber \\
          &= &\sum_\kk w_\k \int \frac{g_\k(R\cos\phi, R\sin\phi, v_z)}{\cos^2 \phi+\sin^2\phi/(1-e_a)^2} dv_z d\phi, \label{eq:sigma}
\end{eqnarray}
where $R$ is the projected distance from the origin to the ellipse defining the isopleth at an angle $\phi$ 
relative to the $x-$axis. In Equation \ref{eq:sigma}, $\Sigma(R_{e})$ is 
free to be scaled by an arbitrary constant factor to minimize $\chi ^2$.
For some systems (e.g., Draco), the available measured surface number density profile is averaged along concentric circles rather than isopleths, we need to simply set $e_a=0$ in Equation (\ref{eq:sigma}).

\section{Simulations}
\label{sec:sim}
To test the algorithm in \S\ref{sec:method}, we construct a dSph galaxy in a power-law potential
with
\begin{eqnarray}
r_0=0.5 \, \hbox{kpc}, \;
\theta=\frac{1}{4}\pi=0.7854, \nonumber\\
\ve{X}=\{\rho_0, \alpha\}=\{8.5 \rhods , 1.2\}, \nonumber\\
f(E, L, L_{z'})= \left(e^{-(E-700)^2/200^2}+0.06 e^{-(E-1650)^2/200^2}\right)e^{sL/L_c(E)}\left(\frac{L_{z'}}{L}\right)^{10}, \label{eq:testDF}
\end{eqnarray}
where $s$ is an anisotropy parameter and $E$ is in units of (\kms)$^2$.
The DF is chosen so that the energy distribution is bimodal to challenge our algorithm.

\begin{figure}
\plottwo{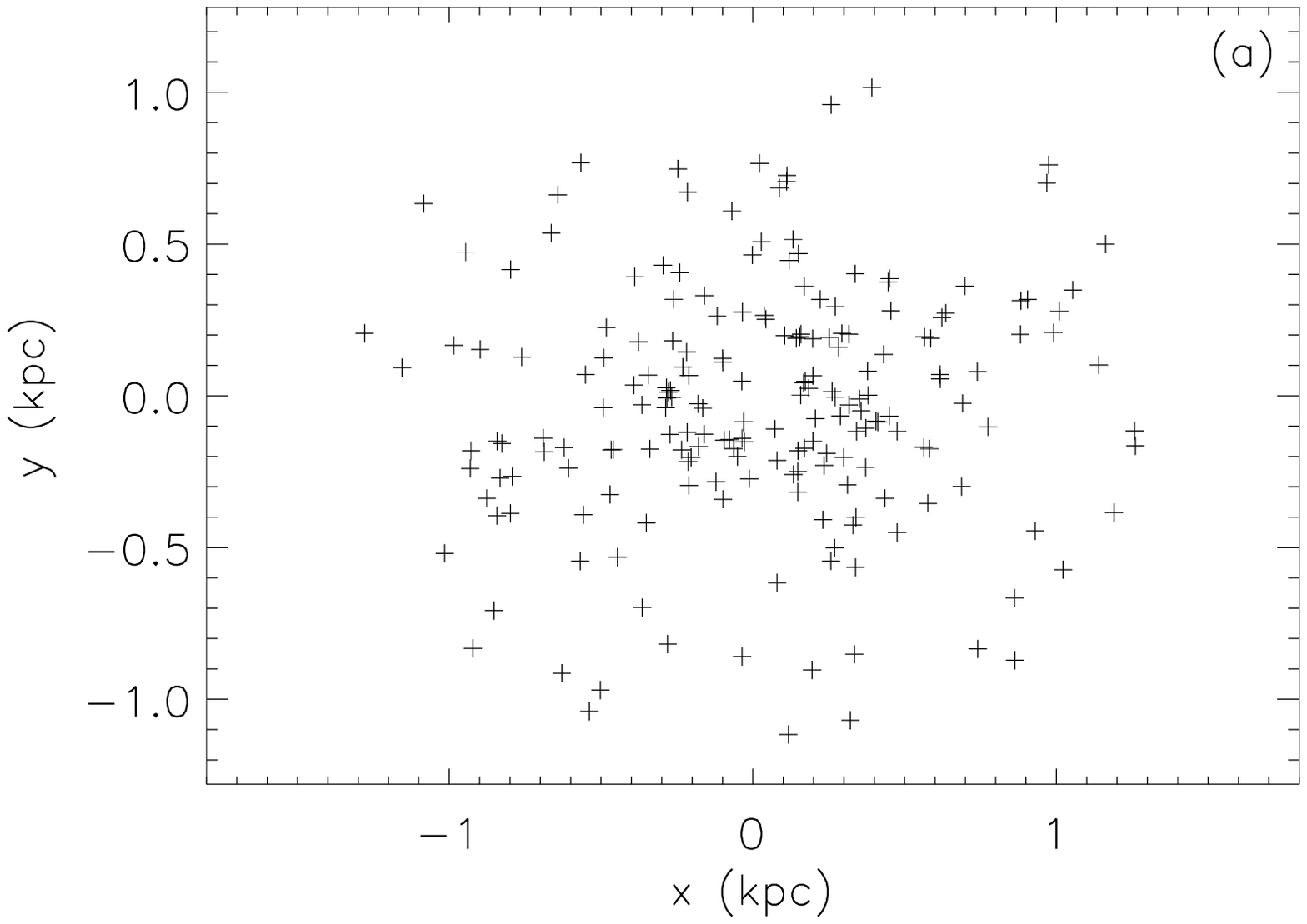}{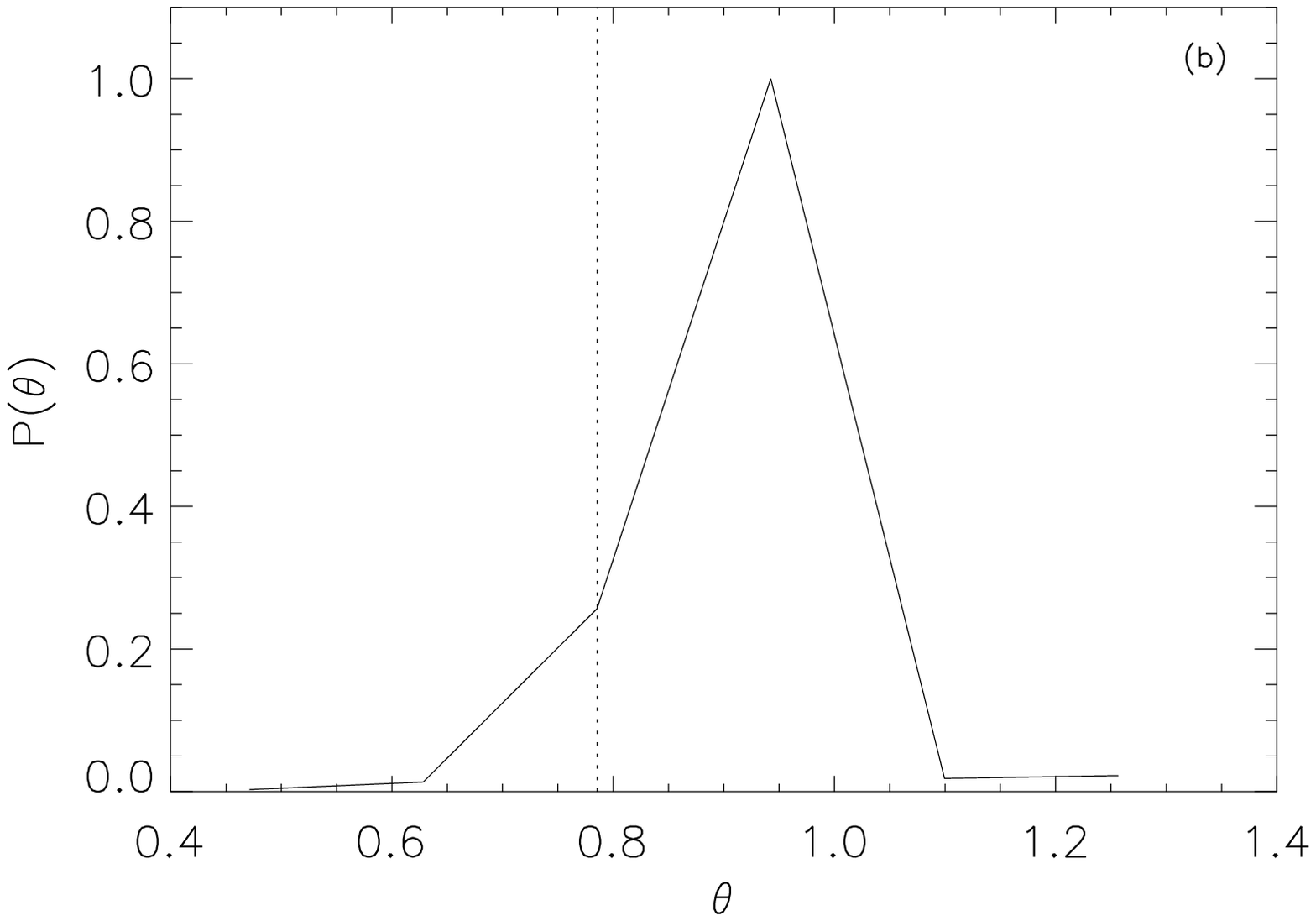}
\plottwo{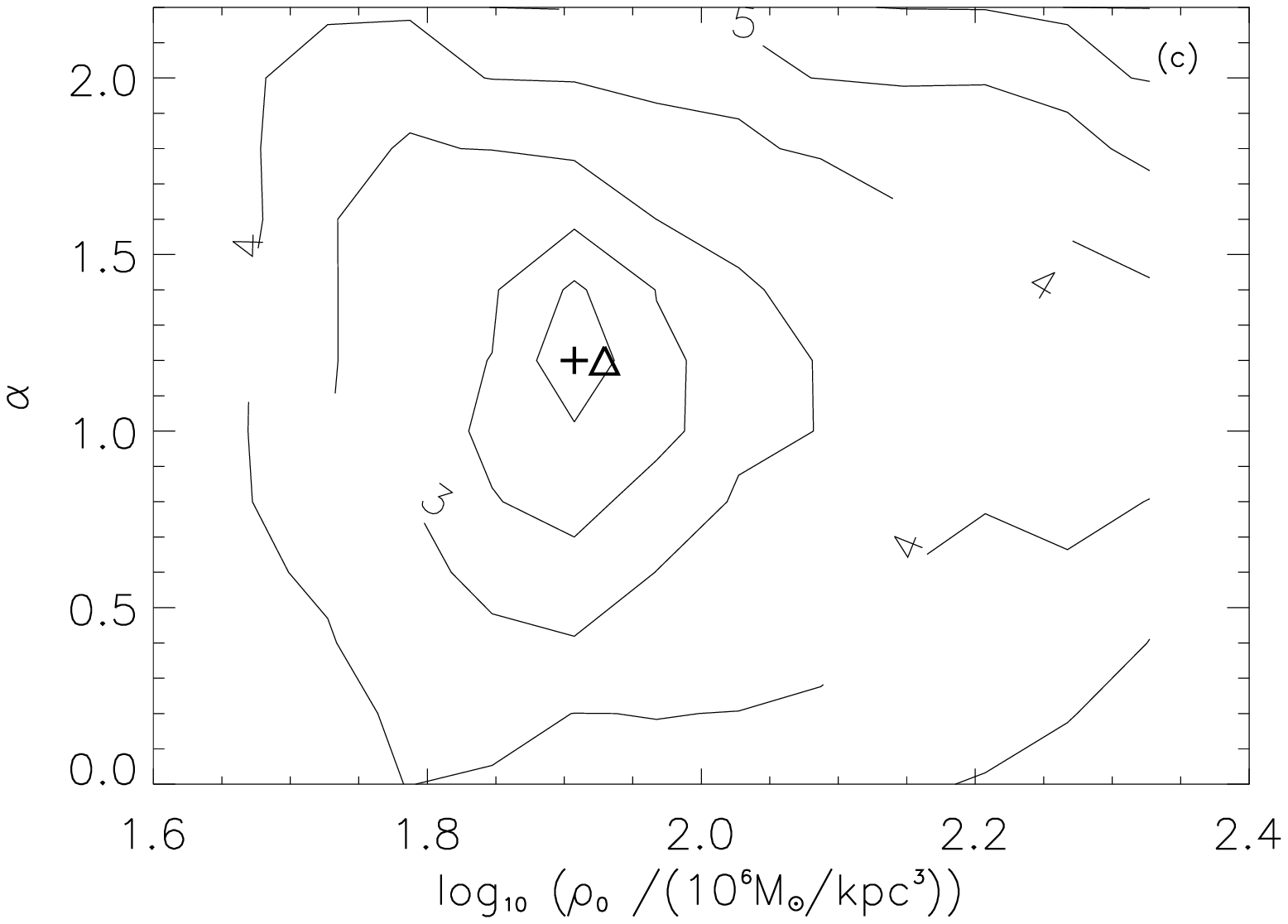}{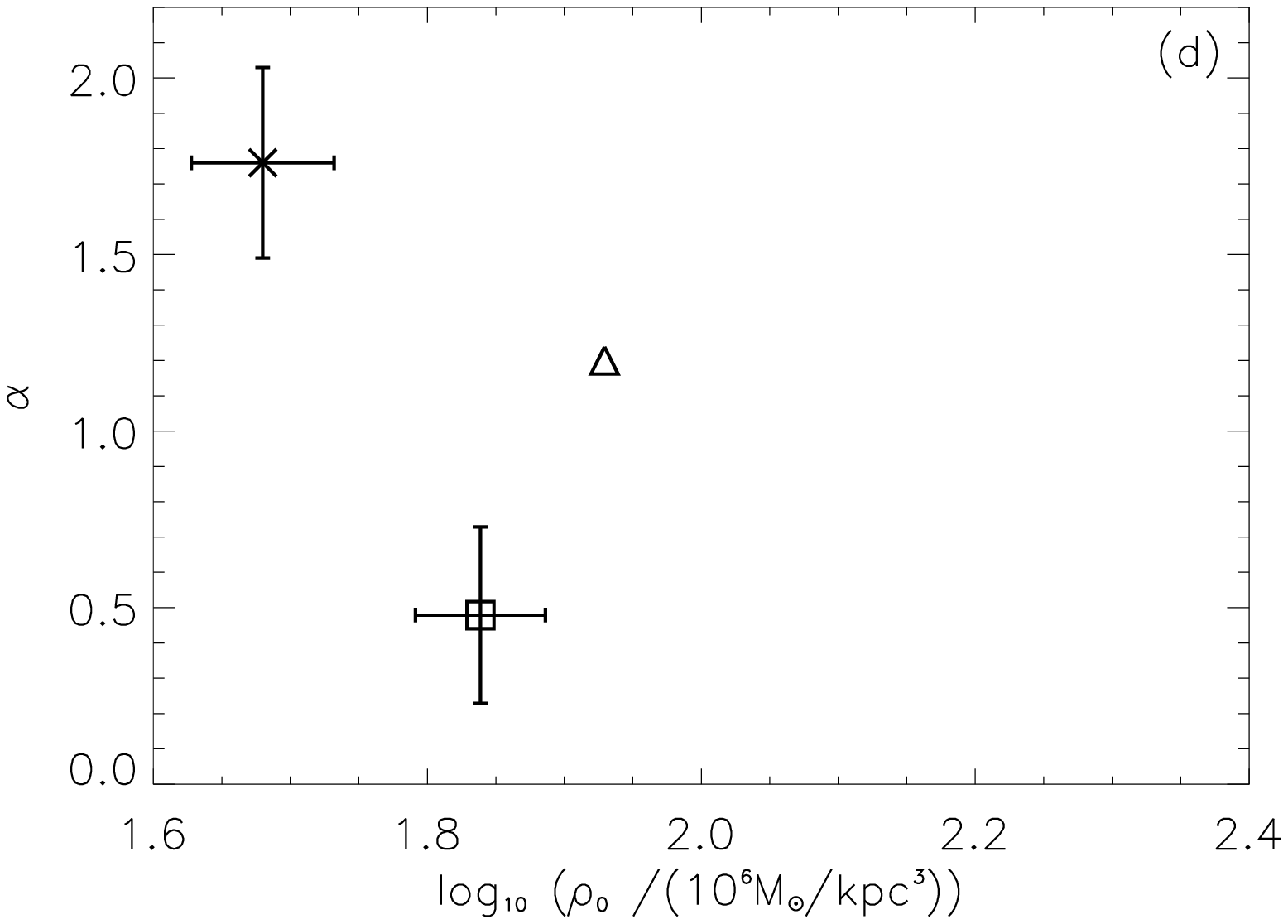}
\figcaption{Algorithm tests.
(a) A test dSph galaxy with 200 observed stars, described by Equation (\ref{eq:testDF}) with $s=5$ (tangential anisotropy). 
(b) Probability distribution of $\theta$ derived from the 200 stars. The input value
is marked by the dotted line. 
(c) Likelihood contours for the potential parameters.
The contours are $n\sigma$ levels from the peak, i.e., $e^{-n^2/2}$ of the maximum likelihood.
The triangle is the input model and the plus sign is the best-fit model with $(N_E, N_L, N_{z'})=(15, 5, 15)$.
They agree within $1\sigma$ error. 
(d) This panel shows the results of fitting flattened anisotropic galaxies
with spherical, isotropic models. The triangle is the input model which is the same model described in (a)
except that $s$ can be $+5$ or $-5$ (radial anisotropy). The square with error bars is the best fit for the model with $s=5$, 
assuming that the stellar part is spherical and the DF is isotropic, $(N_E, N_L, N_{z'})=(15, 1, 1)$, while the
cross is the same as the square except for the input model with $s=-5$.
The large errors in the recovered potential parameters show that both
anisotropy of the DF and non-spherical geometry of the stellar population should be accounted for whenever their effects
may be significant. 
\label{fig:fig1}}
\end{figure}

Figure \ref{fig:fig1}a shows the spatial distribution of 200 stars in the simulated galaxy with $s=5$ (tangential anisotropy),
which has $e_a\sim0.3$. 
We use the algorithm in \S\ref{sec:method} with $(N_E, N_L, N_{z'})=(15, 5, 15)$ and no additional surface density data. 
Accounting for the prior probability of $\theta$ (eq. \ref{eq:ptheta}), 
$\theta$ is derived to be $0.90\pm0.07$ by fitting a Gaussian to the probability distribution in
Figure \ref{fig:fig1}b, which is consistent with the input value at the $1.6\sigma$ level. 
Figure \ref{fig:fig1}c shows the probability distribution of potential parameters, 
specifically, $\ve{X}=\{(8.1\pm1.2)\rhods, 1.17\pm0.25\}$.
The triangle and plus sign are the input model and the best-fit model, respectively, which
agree within the $1\sigma$ error. Therefore, our algorithm can recover both potential parameters
and inclination. 

Contrastingly, with the assumption that the stellar density is spherical and the DF is isotropic, i.e.,
$(N_E, N_L, N_{z'})=(15, 1, 1)$, we may get large systematic errors. We have explored
two cases.
For the above input model with $s=5$ (tangential anisotropy), the assumption leads to the best-fit model marked by the 
square with error bars in panel (d) of Figure \ref{fig:fig1}. For another input model with 
$s=-5$ (radial anisotropy) and the same input potential ${\rho_0, \alpha}$, the best-fit model marked by the cross is also inconsistent with the
input model. We conclude that three-integral models that allow for both velocity anisotropy and flattening must be used to avoid possible systematic errors in the derived potential parameters.

For a face-on galaxy ($e_a=0)$, we find that the $2\sigma$ contour for potential parameters
extends over at least two orders of magnitude in $\rho_0$. 
Because the line-of-sight velocity dispersion is 
decoupled from the mass distribution for a face-on disk-like galaxy,
an apparently round galaxy
may be either a low-density spherical galaxy or a face-on high-density flat galaxy.
Therefore, we get a poor constraint on potential parameters of an apparent round dSph galaxy.
 In practice, we never
  see disk-like dSph galaxies, so when modeling a galaxy with $e_a=0$ we will not make a serious error by
  assuming that the stellar distribution is spherical and using the
  analogous methods described in Paper I for spherical
  systems of test particles\footnote{On the other hand, for a flat edge-on galaxy
($e_a=1.0$), the intrinsic ellipticity $e_i$ is also 1.0, so
we may overestimate the mass by up to $50\%$ if assuming that the stellar distribution is spherical ($f=f(E, L)$).}.
This is not an issue for the three galaxies analyzed in this paper, which are
sufficiently flattened that they yield well-constrained results for the potential
parameters and inclination.

\section{Application to Draco, Ursa Minor and Fornax}
\label{sec:res}
\subsection{Data}
\label{sec:data}
Our algorithm is applied to kinematic data $\{x, y, v_z\}$ and number density profiles $\Sigma_{obs}(R_e)$ of
three dSph galaxies taken from the published literature. 
Table \ref{tab:tab1} lists several parameters of
  each galaxy, including the assumed heliocentric distance ($D$), the
  luminosity in the V-band ($L_V$), apparent ellipticity ($e_a$), and the position angle of the
  major axis derived from the isopleths, as usual measured eastward
  from north ($PA$). In the kinematic sample, we include $N_k$ stars, which have
  heliocentric velocities in the range $v_k\pm\Delta v$ and
  are inside an ellipse with semi-major axis $R_k$ and ellipticity $e_a$, to exclude possible interlopers which could
  strongly bias the results (the same cuts are applied to our
  likelihood distribution, so this procedure introduces no systematic
  errors).
Table 1 also lists the number of stars ($N_s$)
and annuli ($N_a$) in the surface number density profile $\Sigma_{obs}(R_e)$, which has a maximum
radius $R_m$. 
For the outer parts ($R>R_g$), where the number density is very low, we combine a few annuli 
to increase the signal-to-noise ratio. For Draco, we exclude two bins in which the number density is more than 3$\sigma$ away from those in nearby bins. 

\subsection{King model}
\label{sec:king}
\begin{figure}
\plotone{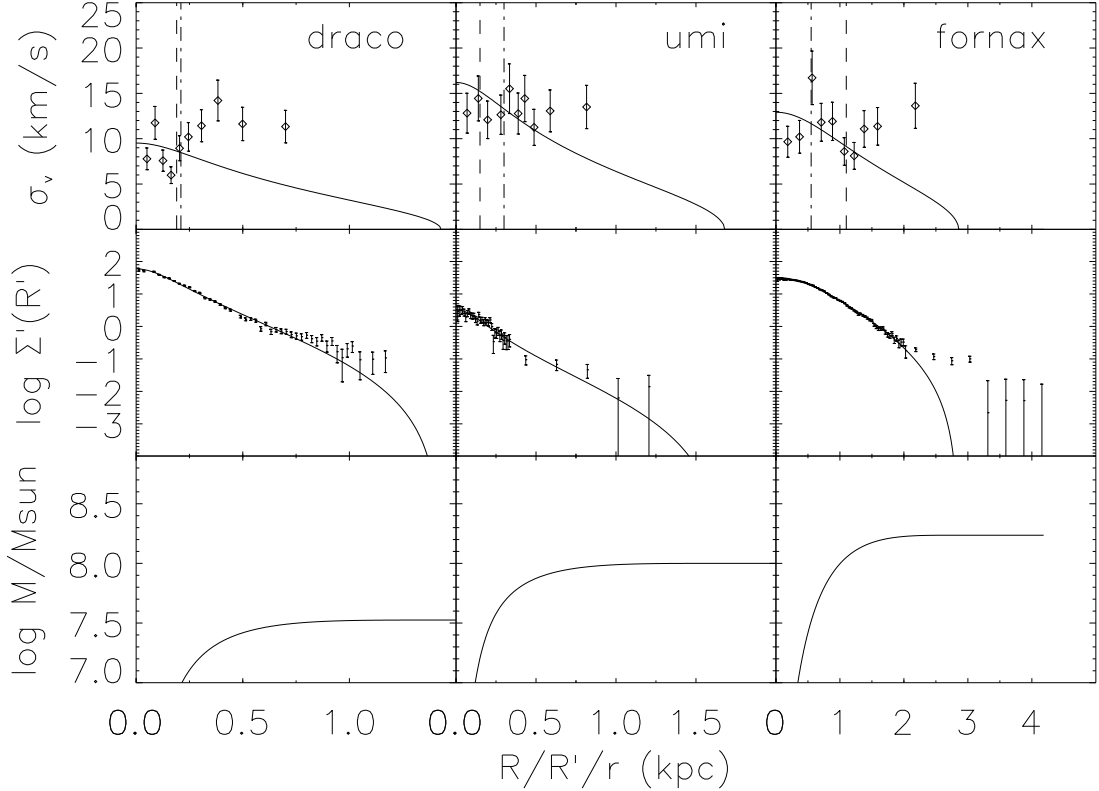}
\figcaption{Fitting King models to dSphs. The upper row shows the observed and best-fit
velocity dispersion profiles $\sigma_{v,obs}$ (points with error bars) and
$\sigma_{v,fit}$ (solid lines). Also shown are the derived core radii (dashed lines)
and those from \citet{IH95} (dash-dotted lines).
The middle row shows the observed density profiles ($\Sigma'(R')$, eq. \ref{eq:R'}) and the best-fit.
The bottom row shows the mass profiles of the best-fit King models. The core radius $r_c$ is 0.19 kpc for
Draco, 0.15 kpc for Ursa Minor and 1.10 kpc for Fornax. A single-component
King model does not fit the data of Draco and Fornax, but is consistent with those of Ursa Minor. Howeve, in all three galaxies the model shows a decline of velocity dispersion with
radius that is not seen in the data.
\label{fig:king}}
\end{figure}

Since King models are spherical, we convert the elliptical surface density
distribution to a spherical one, with a surface number density profile
\begin{eqnarray}
\Sigma'(R')=\Sigma_{obs}(R_e), R'=\sqrt{1-e_a}R_e, \label{eq:R'}
\end{eqnarray}
so that the number of stars enclosed in a circle with radius $R'$ and an ellipse with semi-major
axis $R$ are equal. 
A King model is defined by three parameters, which may be chosen as the characteristic velocity dispersion of 
the DF $\sigma$, the difference between the potential at the tidal radius and that 
at the center $\ve{\Phi}(0)$ in units of $\sigma^2$, 
and a core radius $r_c$ \citep[see][for details]{Bin87}.
Since the DF is parametric, we do not use the method in \S\ref{sec:method}, but
fit velocity dispersion profiles and number density profiles, as shown in 
Figure \ref{fig:king}. We have binned the kinematic data into $N_b=10$ bins with
approximately the same number of stars in each bin. 
Considering the likelihood and assigning a flat prior probability for  $\{\ve{\Phi}(0)/\sigma^2, \ln\sigma\}$,
we estimate the parameters and their errors, which are given in Table \ref{tab:king}.

We fit the observed line-of-sight velocity dispersion profiles while \citet{IH95},
having a smaller kinematic data set, only fit the core
radial velocity dispersion. 
Moreover, \citet{IH95} estimate an average velocity dispersion
$7.5\pm1.0$ \kms from 46 stars for Ursa Minor,
which is much lower than $13.3\pm0.4$ \kms from 163 stars used in our calculation.
Nonetheless, considering the factor of 2.0 in the velocity dispersion
for Ursa Minor, the derive mass-to-light ratios
(see Table \ref{tab:king}) similar to the values of \citet{IH95},
\gml$=245\pm155$ \ml for Draco, $95\pm43$ \ml for Ursa Minor and $7\pm3$ \ml for Fornax.

Figure \ref{fig:king} shows that a single-component
King model does not fit the data of Draco and Fornax, but is formally consistent with those of Ursa Minor.
However, the last point of the Ursa Minor kinematic data suggests that a King model may be ruled out
if future observations show that the flat dispersion profile extends to larger radii. 
In all three systems, the observational data have an excess in
both velocity dispersion and surface number density at large radii, and
the dispersion profile is flat rather than decreasing as in the King models.

\subsection{More general models} 
\label{sec:ml}

We fit the parametrized models for the mass distribution in \S\ref{sec:method} to the data.
We must assume a maximum radius $R_{max}$, so that the upper energy limit of the bins in Equation
(\ref{eq:Ebin}) is $\Phi(R_{max})$. We set $R_{max}$ as 2 kpc for Draco and Ursa Minor, 
6 kpc for Fornax, which are reasonable because the surface number density is 
negligible at these radii. For the power-law model, 
we set $r_0$ as 0.28 kpc for Draco, 0.6 kpc for Ursa Minor and
1.1 kpc for Fornax to minimize the covariance of $\rho_0$ and $\alpha$
in their bivariate probability distribution.

The models labeled ``constant mass-to-light ratio" require some explanation.
Since our method assumes that the potential is
spherical in this paper, we do not attempt to construct a full model of
a non-spherical galaxy with constant mass-to-light ratio. 
To derive the potential of the constant mass-to-light-ratio model, 
we assume that all stars have the same mass-to-light ratio and
construct a spherical density distribution as 
in Equation (\ref{eq:R'}).
So the luminosity profile $\nu_s(r)$ is proportional to the volume number density of stars. 
The spherical assumption is then used to derive the potential, but dropped
when modeling the galaxy itself, i.e., we assume a non-spherical galaxy in a spherical potential.

It is convenient to fit luminosity profile $\nu_s$ 
with a linear combination of two exponential functions  
\begin{eqnarray}
\nu _s(r)=\left\{\begin{array}{ll}
	ae^{-br}+ce^{-dr} &, r<R_{max} \\
	0 & ,r\geq R_{max},
	   \end{array}\right.
\label{eq:fitrou}
\end{eqnarray}
where $a,b,c,d$ are constant parameters listed in Table \ref{tab:exp} so that
\begin{eqnarray}
L_V=4\pi\int _0 ^{R_{max}} \nu_s(r) r^2 dr.
\end{eqnarray}
The chi-square of the fitting shows that the luminosity profiles are indeed 
well approximated by Equation (\ref{eq:fitrou}).

This density distribution gives a potential
\begin{eqnarray}
\Phi(r)=-4\pi G \bar\Gamma_V \left ( \frac {a(2/r-e^{-br}(b+2/r)) }{b^3}+
                                          \frac {c({2}/{r}-e^{-dr}(d+{2}/{r})) }{d^3} \right ). 
\end{eqnarray}
The potential is dependent on $\ve{X}=\{\bar\Gamma_V\}$, the constant mass-to-light ratio in the V-band.

We consider two types of models, those with $N_L=1$, and $N_E$, $N_{z'} \gg 1$, which
correspond to models having $f=f(E, L_{z'})$ (two-integral models), and those with
all of $N_L,N_E$, $N_{z'} \gg 1$ (three-integral models). The two-integral models
are much faster to compute and appear to fit the data just as well as the three-integral models
for the three dSphs, so we focus on the two-integral models.

\subsection{Results}

\label{sec:result}
\subsubsection{Two integral models ($N_L=1$)}
Table \ref{tab:iso} lists the results for various models with $(N_E, N_L, N_{z'})=(15, 1, 20)$, 
the difference in log likelihood $\Delta LH$,  
potential parameters $\ve{X}$, mass and mass-to-light ratio
within $r=R_k$ (see Table \ref{tab:tab1}).
To determine the uncertainties in the derived parameters, we lay down points with
a uniform probability distribution in $\ln(\ve{X})$ if $\ve{X}$ is dimensional and in $\ve{X}$ otherwise, 
then reject or save each point according to the likelihood at that point.
From the saved points, we may estimate the model parameters and their errors. 
In practice, we consider a set of discrete values for $\theta$, 
\begin{eqnarray}
\theta=\frac{i\pi}{20},\;\; \frac{20}{\pi}\arccos(e_a)<i\leq10.
\end{eqnarray}
where $i$ is a natural number.
For all models of the three dwarf spheroidals, either $i=9$ or 10 gives the maximum $P(\ve{X}, \theta)$ and
others give much smaller likelihoods ($3\sigma$ away). Therefore, we do not
try to fit $\theta$ accurately, but give a rough estimate $\theta\sim1.45\pm0.12$.

\begin{figure}
\plotone{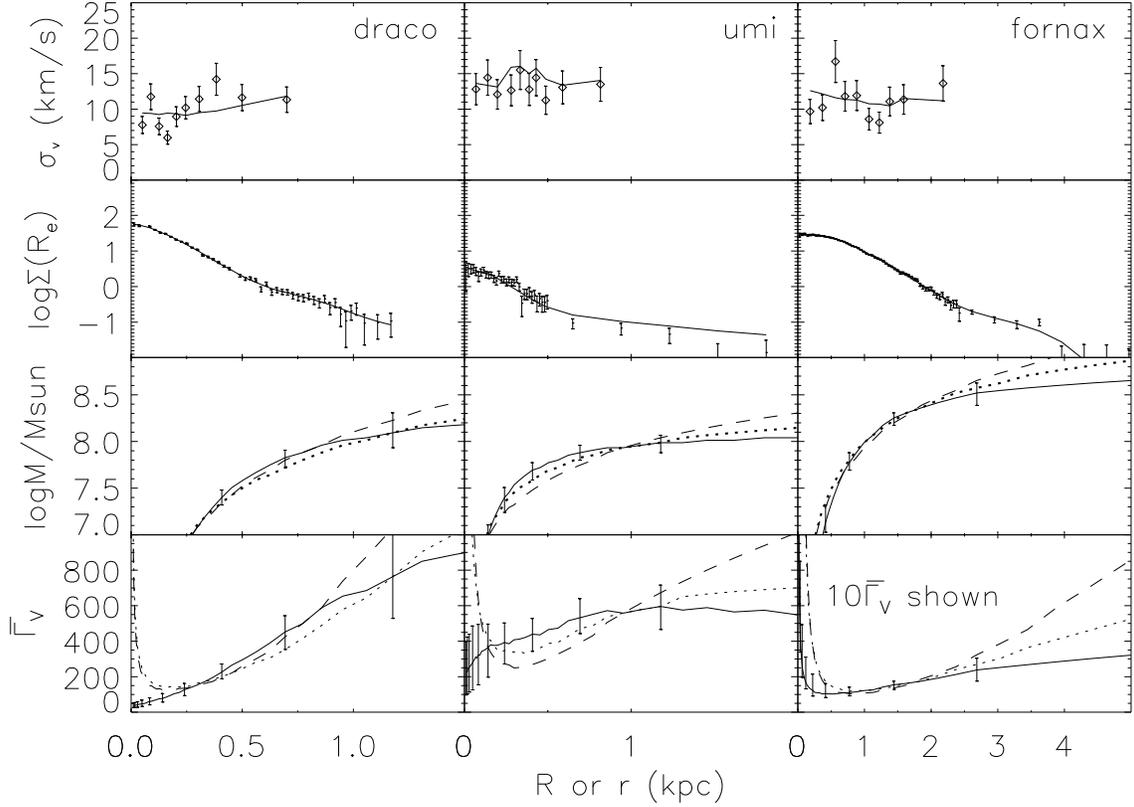}
\figcaption{Best-fit two integral models ($N_L=1$) for three dSph galaxies.
Top row: observed velocity dispersion profiles $\sigma_{v,obs}(R)$ and those of the best-fit models $\sigma_{v,fit}(R)$ (see eq. \ref{eq:h}). Also shown are the core radius $r_c$ (dotted lines) and kinematic survey radius $R_k$ (dashed lines).
Second row: observed surface number density profiles $\Sigma_{obs}(R_e)$ and those of the best-fit models.
Third row: derived mass profiles for the isochrone (solid lines), NFW (dotted lines)
and power-law models (dashed lines).
All three models agree relatively well at intermediate radii.
Bottom row: derived average mass-to-light ratio within radius $r$.
For Fornax, \gml has been multiplied by a constant 10.0 for clear presentation.
The mass-to-light ratios tend to increase with $r$ except at small radii, where
the mass density is very uncertain.
\label{fig:fit}}
\end{figure}

\begin{figure}
\plotone{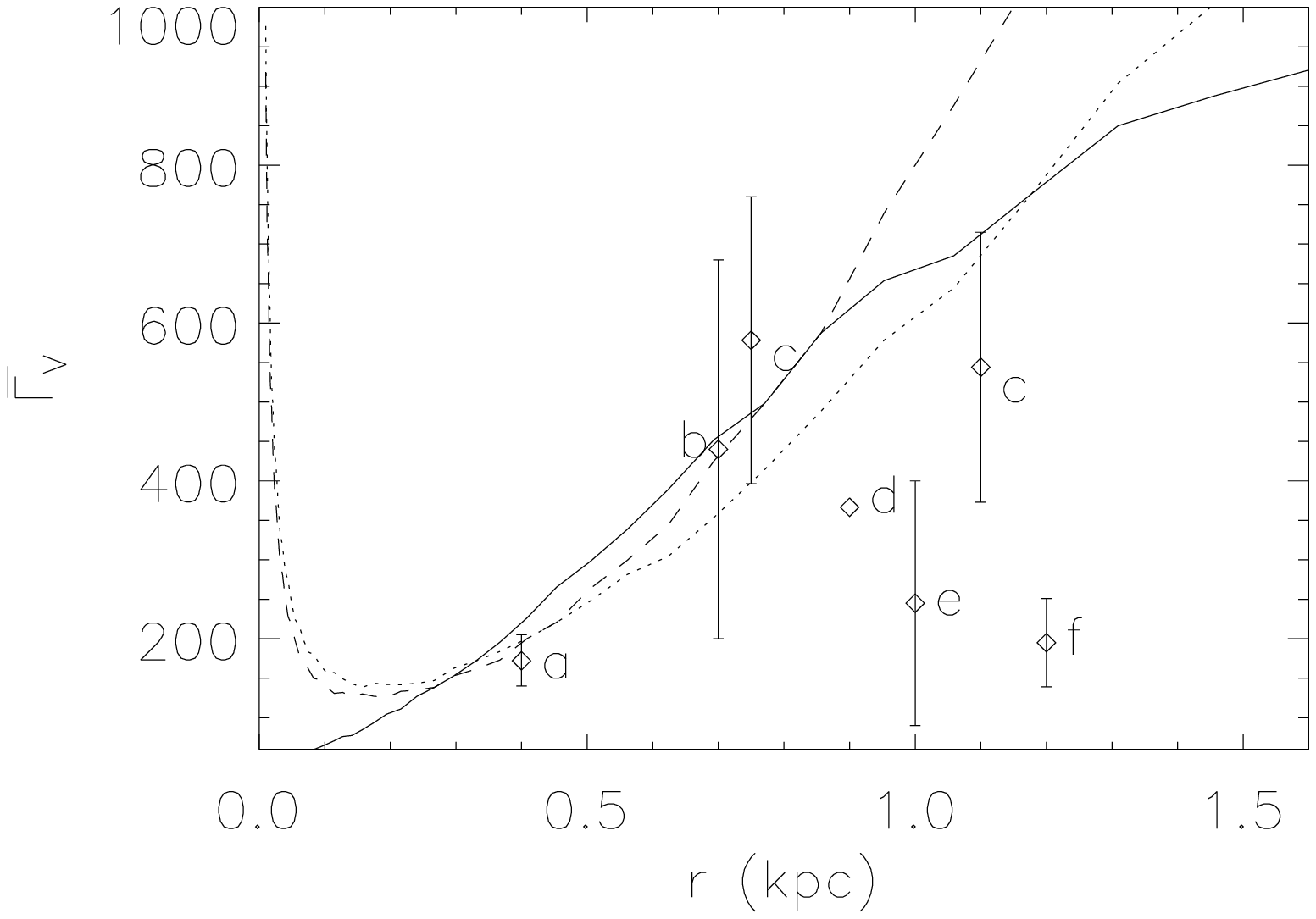}
\figcaption{Derived mass-to-light ratio for the best-fit isochrone (solid line), NFW (dotted line)
and power-law (dashed line) models for two-integral models ($N_L=1$) for Draco.
Also plotted are measurements by a: \citet{Lok02} b: \citet{Kle02} c: \citet{Mas04} d: \citet{Lok05}
e: \citet{IH95}, f: \citet{Ode01}. The average mass-to-light ratio $r$ is assumed constant over $r$ in references $d$ and $e$ but is an estimate at
the radii shown for other measurements.
\label{fig:draco}}
\end{figure}

       Several features of Table 4 are worth noting: (i) For the NFW
        and isochrone models, there is a strong correlation between
        the two parameters in $\ve{X}$, which explains why the
        fractional error in $M(r<R_k)$ is smaller than the fractional
        error in the potential parameters. (ii)
        The best-fit isochrone, NFW and power-law models have $-4.2\leq\Delta
        LH<0$, corresponding to $<2.9\sigma$. The models of a dominant central black hole,
        constant density and (for Draco and Fornax) constant mass-to-light ratio are less
        likely at $>3.7\sigma$ levels.
        (iii) The difference in
        log likelihood only indicates the relative likelihood among
        different models, so it is important to investigate whether
        the best-fit model is a good fit. In Figure \ref{fig:fit}
        we show the observed velocity dispersion $\sigma_{v,obs}(R)$
        in ten bins (top row). For the stars in a bin centered at $R$, say, $\{x_j, y_j, v_{zj}\}$,
        we calculate the predicted velocity dispersion $\sigma_{v,fit}(R)$
        from the velocity distribution profile
        \begin{eqnarray}
        h(v_z|R)=\sum_j g(x_j, y_j, v_z). \label{eq:h}
        \end{eqnarray}
        Comparison of the first two
        rows of panels in Figures \ref{fig:king} and \ref{fig:fit} shows that the present
        models provide a substantially better fit to the observed
        velocity-dispersion and surface-density profiles in all three
        galaxies than the best-fitting King models. This visual
        impression is confirmed in Table 6, which gives
        goodness-of-fit statistics for the velocity-dispersion profile
        ($\chi_v^2$) and surface-density profile ($\chi^2$) and their
        deviations from the expected values $N_b$ and $N_a$ (the
        number of bins), in units of the standard deviation
        $\sqrt{2N_b}$ or $\sqrt{2N_a}$ ($1\sigma$ error). The
        deviations for the present models are $<2.1\sigma$ for all
        three galaxies, while the King models are much poorer fits for
        Draco and Fornax. (iv) Note that the models are fitted to the
        surface-density distribution as a function of radius, assuming
        the observed ellipticity $e_a$ (eqs. 14-17). Thus it is not
        guaranteed that the the ellipticity of the isopleths in the
        models will be similar to the observed ellipticity (no doubt
        a better approach would have been to fit the two-dimensional
        surface-density distribution, but this is substantially more
        computationally expensive). Nevertheless the ellipticities of
        the models, quoted in Table 6 at the core radius, are similar
        to the observed ellipticities.

The third row in Figure \ref{fig:fit} shows the derived mass profiles of various models for the three dSph galaxies.
All of the models shown are within $2.5\sigma$ of the best-fit models. The models have similar density distributions
at intermediate radii, for example, $0.25-1.0$ kpc for Draco, but (not surprisingly) differ substantially
at radii beyond the outermost observational data.
To summarize, our algorithm gives relatively small errors at intermediate
radii, and the surface number density profile only weakly constrains the mass density profile
beyond $r=R_k$. 

The bottom row in Figure \ref{fig:fit} shows the average mass-to-light ratio \gml$(r)$ for the
best-fit models. The light inside a radius $r$ is calculated from Equation (\ref{eq:fitrou}). 
Comparing Table \ref{tab:king} and \ref{tab:iso},
we can see that King model gives estimates of mass at the outer limit of the kinematic data,
$r=R_k$, consistent with those
from more general models within $2\sigma$. However, King model assumes a constant
mass-to-light ratio, which is an average value.
For the more general models, except at small radii where the
mass density is very uncertain, the mass-to-light ratios increase with radius.
Therefore, the global mass-to-light ratios are probably even greater than \gml$(R_k)$.

\subsubsection{Three integral models ($N_L=5$)}
We have shown that two integral ($N_L=1$) models can reproduce all of the observed features of dSph galaxies.
We also list the results for three-integral models, $(N_E, N_L, N_{z'})=(15, 5, 20)$, 
in Tables \ref{tab:aniso} and \ref{tab:stat}.
The mass-to-light ratios are generally larger than the values for $N_L=1$ models by $10\%$ to $50\%$. 
This result suggests that given the freedom to adjust the DF in $L$ direction, high mass
models are better able to fit the observed data than low mass models, thus the
derived mass distributions are more skewed to the high end. 
In the following sections, we adopt the conservative (lower) estimates of mass-to-light ratios that are derived with the case of $N_L=1$.

\section{Discussion}
\label{sec:dis}

We believe that the mass estimates presented here are the most accurate and reliable that have
so far been obtained for the three dSph galaxies, Draco, Ursa Minor and Fornax. 
In our models, we assume that the stellar distributions are not significantly affected by the tidal force of the Milky Way.
This assumption is confirmed by a dynamical argument: using the mass $M(r<R_k)$ of the best-fit models in Table \ref{tab:iso}, which is a lower limit to the total mass,
the Roche limit is 2.6 kpc (3.5$R_k$) for Draco, 2.6 kpc (2.4$R_k$)
for Ursa Minor and 6.6 kpc (2.4$R_k$) for Fornax, assuming that that the Milky Way is an isothermal 
sphere with $v_{circular}=220$ \kms. The Roche limits also exceed $R_{max}$, the maximum radius
used in our dynamic models in 
\S\ref{sec:ml}.
Therefore, our assumption that the dSph galaxies are isolated systems
in virial equilibrium is self-consistent.
Because our models can fit both radial velocity profiles and surface number density profiles, the so-called
``extra-tidal extensions'' in the surface number density profiles found by \citet{IH95} and \citet{Wil04} 
do not require any special ad hoc explanation. Thus it is not
valid to consider them as the evidence of tidal stripping, as proposed by some authors 
\citep[e.g.,][]{Mar01, Go03, Mun05}. 

We strongly rule out any centrally condensed mass distribution like a dominant central black hole, as well
as constant-density and (for Draco and Fornax) constant mass-to-light ratio models.
For example, the constant-density model is ruled out at the level of 4.1$\sigma$ for Draco, 
3.4$\sigma$ for Ursa Minor and 4.4$\sigma$ for Fornax. In contrast to the positive result 
of \citet{Lok05} based on the same data set, the constant mass-to-light ratio is ruled out at 8.6$\sigma$ level for Draco.

We derive an average mass-to-light ratio of $400\pm80$ \ml
within $0.75$ kpc for Draco,
$580^{+140}_{-100}$ \ml within $1.1$ kpc for Ursa Minor and
$25^{+7}_{-5}$ \ml within $2.5$ kpc for Fornax.
Note that these values are probably lower limits for the global mass-to-light ratio 
because the mass-to-light ratio increases with $r$ for the best-fit models.
The mass-to-light ratios we obtain for all three galaxies are larger than the estimates of global mass-to-light ratio by \citet{IH95}
($245\pm155$ \ml for Draco, 
$95\pm43$ \ml for Ursa Minor and $7\pm3$ \ml for Fornax). 
The discrepancies arise because the results of \citet{IH95} and \citet{Ode01} are 
based on a single-component King model and because we have
a larger sample of radial velocities, covering a larger range of radii.

Figure \ref{fig:draco} shows the derived mass-to-light ratio of Draco as a function of
radius and the estimates by
other authors. 
The mass-to-light ratio for Draco is assumed constant in some
of these studies \citep{IH95,  Lok05}
but is allowed to vary with radius in others \citep{Ode01, Lok02, Kle02, Mas04}.
\citet{Pia02} have proposed that
  the large mass-to-light ratios of dSph's may be partly due to
  velocity anisotropy, since anisotropy can inflate the derived \gml by
  a factor of three or so if the kinematic information is restricted
  to the central parts of the galaxy.
The simulations of \S\ref{sec:sim} confirm that velocity anisotropy and flattening
  of the stellar distribution (i.e., the dependence of the DF on $L$
  and $L_{z'}$) must be accounted for to obtain accurate mass-to-light
  ratios.
Nonetheless, for the three dSph galaxies, the DF $f(E, L_{z'}/L)$ fits the data reasonably well 
(\S\ref{sec:result}), and the derived masses
actually increase if we allow the dependence of DF on $L$ (\S\ref{sec:ml}).
So we find no evidence that three-integral models fit the current data significantly
better than two-integral models, or that they prefer models with lower mass-to-light ratio.

\citet{Wil04} and \citet{Kleyna04} argue that
  the velocity dispersion profiles of Draco and Ursa Minor seem to
  show a sharp drop at radii of 0.81 and 0.84 kpc, respectively. The drop in
  the dispersion profile seems to coincide with a ``break'' in the
  number density profile, and Wilkinson et al. propose that this is
  evidence of a kinematically cold population. However, the sharp drop
  may only be present in particular binning schemes \citep{Mun05}
  and is not seen with our binning scheme (Figure \ref{fig:king}). Our
  models do not reproduce such a sharp drop.
  Thus, the drop is inconsistent with stationary axisymmetric
  models of the DF and also is not statistically significant given the
  present data.

The large masses that we have derived may help alleviate but
may not eliminate the problem that there are far fewer dwarf galaxies in the Local Group
than predicted by standard cosmological models.

\section{Conclusion}
\label{sec:con}
We model dSph galaxies as axisymmetric stellar systems in spherical potentials, which are in 
dynamical equilibrium without significant external tidal forces. 
For the three dSph galaxies, Draco, Ursa Minor and Fornax, plausible parametric
  models of the potential combined with non-parametric two-integral
  models of the stellar distribution function can
naturally reproduce the observed velocity dispersion profiles and number density profiles. The masses
of these models are high enough so that tidal forces from the Milky Way do not 
significantly affect the observed kinematics.
So the models are self-consistent and ``extra-tidal extensions'' simply reflect
an extended axisymmetric envelope that is still in virial equilibrium.
The sharp drop in velocity dispersion 
which coincides with a ``break'' in the light distribution that has
been claimed in the literature is not observed
in our binning scheme and can not be reproduced
in the scenario of virial equilibrium. 

The isochrone, NFW and power-law models all fit the data.
We strongly rule out any centrally condensed mass distribution like a dominant central black hole, as well
as constant-density and (for Draco and Fornax) constant mass-to-light ratio. 
The average V-band mass-to-light ratio in our best-fit two-integral models is
$400\pm80$ \ml
within $0.75$ kpc for Draco,
$580^{+140}_{-100}$ \ml within $1.1$ kpc for Ursa Minor and
$25^{+7}_{-5}$ \ml within $2.5$ kpc for Fornax.
Our method considers both anisotropy of the DF and the non-spherical geometry 
of the stellar population, and investigates a large variety of mass distribution models, 
and thus should yield more reliable mass distributions than previous investigations.

Our simulations show that anisotropy
  (dependence on $L$) and flattening (dependence on $L_{z'}$) in the
  DF may substantially affect derived masses of axisymmetric
  galaxies. However, we detect no strong evidence of anisotropy in the
  three dSph galaxies, i.e., a two-integral DF $f(E,L_{z'})$ is
  consistent with the data. Allowing anisotropy yields even larger
  mass-to-light ratios.

Our method could be improved by
  fitting the two-dimensional surface number-density distribution
  rather than a one-dimensional profile. Observationally, it would be
  useful to extend the kinematic surveys to larger radii, since the
  current limits are still well within the radii where tidal forces
  are likely to complicate the analysis.

\acknowledgements
I thank Scott Tremaine for numerous stimulating discussions and helpful
comments on an earlier draft, James Gunn, David Spergel, Michael Strauss and Glenn van de ven for helpful discussions.
I wish to thank Mark Wilkinson for providing the kinematic data of Draco and Ursa Minor 
and surface number density
profile of Draco in tabular form.
This research was supported in part by NASA grant NNG04GL47G and used
computational facilities supported by NSF grant AST-0216105.

\begin{deluxetable}{lcccccc}
\tablecaption{Data summary\tablenotemark{a} \label{tab:tab1}}
\tablehead{
                        & \colhead{Draco} & \colhead{Ursa Minor} & \colhead{Fornax} & \colhead{Reference} }
\startdata
\sidehead{galaxy parameters}
\hline
$D$ (kpc)                  & 82   & 66   &   138   & 1\\
$L_V$ ($10^5L_{V,\odot}$)  & 1.8  & 2.0  &  140     & 1, 2 \\
$e_a$                      & 0.29 & 0.56 &  0.30   & 2\\
$PA$ (degree)              & 88   & 56   &  41    & 2, 3 \\
\hline
\sidehead{kinematic data} 
\hline 
$N_k$          &  197              & 163             &  147          & 4, 5 \\
$v_k\pm\Delta v$ (\kms)   & $-290.7$$\pm$30.0 &$ -245.2\pm$35.0 &  53.3$\pm$30.0 & \nodata \\
$R_k$ (kpc)    & 0.75               & 1.1             &  2.5          & \nodata \\
\hline 
\sidehead{surface number density profiles}
\hline
$N_s$               & 18000  & 1500  & 42000 & 2, 4\\
$R_m$ (kpc)         & 1.2    & 1.3   &  4.0  & \nodata \\ 
$R_g$ (kpc)         & n/a    & 0.5   &  2.5  &\nodata \\
$N_a$               & 44     & 38    &  66   &\nodata \\ 
\enddata
\tablenotetext{a}{See section \S\ref{sec:data} for explanations of the parameters.}
\tablerefs{(1)\citet{Mat98}; (2) \citet{IH95}; (3) \citet{Ode01}; (4) \citet{Wil04}; (5) \citet{Wal06}.}
\end{deluxetable}

\begin{deluxetable}{lccccccc}
\tablecaption{Parameters of King models for dSph galaxies\label{tab:king}}
\tablehead{\colhead{Galaxy} & \colhead{$\frac{\ve{\Phi(0)}}{\sigma^2}$} & \colhead{$\sigma$} & \colhead{$r_c$} & \colhead{$r_t$} & \colhead{$M_{total}$} & \colhead{$M/L$} & \colhead{$M(r<R_k)$} \\
             &  & (\kms)     & (kpc) & (kpc) & ($10^6$M$_\odot$) & (M$_\odot$/L$_\odot$) & ($10^6$M$_\odot$)
}
\startdata
Draco      & $4.3\pm0.1$ & $11\pm1$ & $0.182\pm0.005$ & $1.42\pm0.04$  &$34\pm5$ & $180\pm30$ & $28\pm4$\\
Ursa Minor & $5.0\pm0.6$ & $18\pm2$ & $0.18\pm0.02$ & $1.7\pm0.3$  &$100\pm20$ & $500\pm150$ & $83\pm28$\\
Fornax     & $1.5\pm0.1$ & $23\pm2$ & $1.11\pm0.05$ & $2.84\pm0.04$  &$180\pm20$ & $11\pm4$ & $160\pm50$\\
\enddata
\end{deluxetable} 

\begin{deluxetable}{lcccccccc}
\tablecaption{Parameters of luminosity profiles\label{tab:exp}}
\tablehead{
    \colhead{} & \colhead{$R_{max}$} &\colhead{$a$} & \colhead{$b$} & \colhead{$c$} & \colhead{$d$} &\colhead{$N_a$} & \colhead{$\chi^2$} & \colhead{$|N_a-\chi^2|$}\\
      &  \colhead{(kpc)} & \colhead{(L$_\odot$ kpc$^{-3}$)}    & \colhead{(kpc$^{-1}$)} & \colhead{(L$_\odot$ kpc$^{-3}$)}  & \colhead{(kpc$^{-1}$)} &\colhead{} & \colhead{} & \colhead{($\sqrt{2N_a}$)}
}
\startdata
Draco          & 2.0 &  $4.69\times10^6$ & $9.17$ & $3.26\times10^3$ &  $0.98$ & 44 & 68 & 2.6\\
Ursa Minor     & 2.0 &  $6.27\times10^6$ & $10.4$ & $1.01\times10^4$ &  $1.22$ & 38 & 31 & 0.8\\
Fornax\tablenotemark{a}         & 6.0 &  $-4.21\times10^7$ & $4.79$ & $4.21\times10^7$ &  $3.55$ & 66& 95 & 2.5\\ 
\enddata
\tablenotetext{a}{We perform the optimization with the constraint that the central density
$a+c\geq0$.}
\end{deluxetable}

\begin{deluxetable}{ccccccc}
\tablecaption{Results from two-integral models of dwarf spheroidal galaxies\label{tab:iso}}
\tablehead{
   \colhead{Mass model\tablenotemark{a}}  & \colhead{$\Delta LH$ } & \colhead{$\ve{X}\{1\}$\tablenotemark{b}} &\colhead{$\ve{X}\{2\}$\tablenotemark{c}} & \colhead{$M(r<R_k)$\tablenotemark{d}} & \colhead{$\bar\Gamma_V (r<R_k)$\tablenotemark{e}} \\
      &  & &  & $(10^6M_\odot)$ &(\ml) & }
\startdata
\sidehead{Draco}
\hline
 1 & $-0.3$   &  $180^{+140}_{-50}$     &  $0.48^{+0.23}_{-0.12}$         &  $73^{+24}_{-16}$   & $480^{+160}_{-110}$ \\
 2 & $0.0$    &   $16^{+26}_{-12}$     & $2.0^{+3.3}_{-1.0}$  &  $60\pm12$   & $400\pm80$  \\
 3\tablenotemark{f} & $-0.1$     &  $77\pm8$      & $1.1^{+0.2}_{-0.5}$         &  $72^{+24}_{-16}$ & $480^{+160}_{-110}$   \\
 4 & $-8.0$    &  $80^{+40}_{-20}$  & \nodata         &  $ 77^{+47}_{-27}$        & $510^{+310}_{-180}$     \\
 5 & $-13.7$   &  $420\pm60$    & \nodata             &  $ 63\pm9$          & $420\pm60$         \\
 6 & $-36.8$   &  $38\pm4$      &\nodata             &  $38\pm4$           & $250\pm30$          \\
\hline
\sidehead{Ursa Minor}
\hline
 1 & 0.0       & $140^{+50}_{-30}$  & $0.18^{+0.08}_{-0.04}$        &  $93^{+23}_{-16}$  & $580^{+140}_{-100}$ \\
 2 & $-3.3$   & $970^{+700}_{-600}$ & $0.20^{+0.13}_{-0.05}$   &  $93^{+21}_{-16}$  & $580^{+130}_{-100}$   \\
 3 & $-4.2 $   & $22\pm4$ & $1.8^{+0.2}_{-0.4}$  &  $101^{+13}_{-17}$ & $630^{+80}_{-110}$   \\
 4 & $-20.0$   &  $22^{+11}_{-7}$ & \nodata        &  $100^{+55}_{-36}$  & $630^{+340}_{-230}$  \\
 5 & $-4.0$    &  $720\pm60$ & \nodata        &  $150\pm10$         & $720\pm60$ \\
 6 & $-14.9$   &  $80\pm9$    & \nodata        &  $80\pm9$          & $500\pm60$ \\
\hline
\sidehead{Fornax}
\hline
 1 & 0.0    & $660^{+370}_{-240}$ & $0.9^{+0.2}_{-0.3}$      & $340^{+90}_{-70}$ & $25^{+7}_{-5}$ \\
 2 & $-0.7$ & $13^{+17}_{-7}$ & $2.2^{+1.4}_{-1.0}$    & $350^{+100}_{-90}$         & $25^{+7}_{-6}$       \\
 3 & $-1.3$ & $19\pm2$ & $1.4^{+0.2}_{-0.4}$               & $390^{+110}_{-80}$  & $28^{+9}_{-6}$   \\
 4 & $-9.7$ & $20^{+9}_{-6}$ &\nodata              & $500^{+450}_{-200}$         & $36^{+33}_{-14}$       \\
 5 & $-6.9$ & $17\pm3$ & \nodata                   & $240\pm40$         & $17\pm3$        \\
 6 &$-17.7$ & $180\pm30$ & \nodata                   & $180\pm30$         & $13\pm2$       \\
\enddata

\tablenotetext{a}{Model: 1. isochrone, 2. NFW, 3. power-law, 4. constant density, 5. constant mass-to-light ratio, 6. black hole. See \S\ref{sec:method} for description of the models.
}
\tablenotetext{b}{The first parameter of mass distribution models: mass for isochrone and
black hole models in units of $10^6$ M$_\odot$, density for NFW, power-law, 
and constant density models in units of $10^6$ M$_\odot$\ kpc$^{-3}$, mass-to-light ratio for 
constant mass-to-light ratio model in units of M$_\odot/$L$_{V,\odot}$. See Equations (\ref{eq:iso})-(\ref{eq:bh}) for definitions 
of the parameters.}
\tablenotetext{c}{The second parameter of mass distribution models: radius for isochrone and NFW models in units of kpc, density exponent for power-law model.}
\tablenotetext{d}{Mass interior to to the limiting radius $R_k$ of kinematic data. $R_k=0.75$ kpc for Draco, 1.1 kpc for Ursa Minor and 2.5 kpc for Fornax.}
\tablenotetext{e}{The average mass-to-light ratio within $r=R_k$.}
\tablenotetext{f}{The arbitrary parameter $r_0$ (eq. \ref{eq:powerlaw}) is chosen as 0.28 kpc for Draco, 0.6 kpc for Ursa Minor and 1.1 kpc for Fornax.}

\end{deluxetable}

\begin{deluxetable}{ccccccc}
\tablecaption{Results from three-integral models of dwarf spheroidal galaxies\tablenotemark{a}  \label{tab:aniso}}
\tablehead{
   \colhead{Mass model}  & \colhead{$\Delta LH$ } & \colhead{$\ve{X\{1\}}$} &\colhead{$\ve{X\{2\}}$} & \colhead{$M(r<R_k)$} & \colhead{$\bar\Gamma_V (r<R_k)$} \\
      &  & &  & $(10^6$M$_\odot)$ &(\ml) & }
\startdata
\sidehead{Draco}
\hline
 1 & $-0.3$   &  $330^{+180}_{-160}$     &  $0.51^{+0.08}_{-0.16}$         &  $74^{+19}_{-16}$   & $490^{+130}_{-110}$ \\
 2 & $-0.3$    &   $26^{+25}_{-17}$     & $1.2^{+2.0}_{-0.5}$  &  $58^{+20}_{-13}$   & $380^{+130}_{-70}$  \\
 3 & $0.0$     &  $81\pm14$      & $1.1\pm0.4$         &  $77\pm26$ & $510\pm170$   \\
 4 & $-9.5$    &  $83^{+40}_{-30}$  & \nodata         &  $ 83^{+40}_{-30}$        & $510^{+310}_{-180}$     \\
 5 & $-5.2$   &  $370\pm40$    & \nodata             &  $ 56\pm6$          & $370\pm40$         \\
 6 & $-31.9$   &  $29\pm3$      &\nodata             &  $29\pm3$           & $190\pm20$          \\
\hline
\sidehead{Ursa Minor}
\hline
 1 & $-0.3$       & $170\pm50$  & $0.21^{+0.04}_{-0.08}$   &  $105^{+29}_{-25}$  & $660^{+180}_{-160}$ \\
 2 & $-0.1$   & $300^{+1100}_{-130}$ & $0.37^{+0.12}_{-0.19}$   &  $114^{+28}_{-21}$  & $710^{+180}_{-130}$   \\
 3 & $0.0 $   & $30^{+7}_{-6}$ & $1.4\pm0.3$  &  $125^{+26}_{-22}$ & $780^{+160}_{-130}$   \\
 4 & $-16.4 $   &  $22^{+11}_{-8}$ & \nodata        &  $100^{+56}_{-36}$  & $630^{+350}_{-230}$  \\
 5 & $-1.7$    &  $720\pm60$ & \nodata        &  $115\pm10$         & $720\pm60$ \\
 6 & $-11.1$   &  $88\pm9$    & \nodata        &  $88\pm9$          & $550\pm60$ \\
\hline
\sidehead{Fornax}
\hline
 1 & $-2.5$    & $2100^{+600}_{-500}$ & $1.5^{+0.2}_{-0.3}$      & $520^{+90}_{-100}$ & $38\pm7$ \\
 2 & $-0.1$ & $5.9^{+0.6}_{-0.3}$ & $4.0\pm0.4$    & $460^{+60}_{-80}$         & $33^{+4}_{-6}$       \\
 3 & $0.0$ & $25^{+3}_{-7}$ & $1.3^{+0.2}_{-0.3}$               & $500^{+70}_{-40}$  & $36^{+5}_{-3}$   \\
 4 & $-13.0$ & $22^{+11}_{-8}$ &\nodata              & $550^{+530}_{-250}$         & $40^{+39}_{-18}$       \\
 5 & $-9.7$ & $15\pm2$ & \nodata                   & $210\pm30$         & $15\pm2$        \\
 6 &$-17.8$ & $180\pm40$ & \nodata                   & $180\pm40$         & $13\pm3$       \\
\enddata
\tablenotetext{a}{Definitions and units of all parameters are the same as those in Table \ref{tab:iso}} 
\end{deluxetable}

\begin{deluxetable}{lccccccc}
\tablecaption{Statistics of the best-fit models \label{tab:stat}}
\tablehead{  \colhead{Model} & \colhead{$e_a$} &  \colhead{$N_b$} & \colhead{$\chi_v^2$} & \colhead{$|{\chi_v^2-N_b}|$} & \colhead{$N_a$} & \colhead{$\chi^2$} &\colhead{$|\chi^2-N_a|$} \\
 & & &  & ($\sqrt{2N_b}$) &  &  & ($\sqrt{2N_a}$)
}
\startdata
\sidehead{Draco}
\hline
King    &  0.0  & 10 & 54.3  & 9.9 & 44 & 108.5 & 6.9      \\
$f(E, L_{z'})$ &  0.25  & 10 & 21.1  & 2.5 & 44 & 38.7 & 0.6   \\
$f(E, L, L_{z'})$ &  0.30  & 10 & 21.0  & 2.4 & 44 & 32.2  & 1.3  \\
\hline
\sidehead{Ursa Minor}
\hline
King  & 0.0  & 10  & 13.9& 0.9 & 38 & 40.3 &  0.3          \\
$f(E, L_{z'})$ & 0.35 & 10  & 6.3 & 0.8 & 38 & 53.2 & 1.7 \\
$f(E, L, L_{z'})$ & 0.40 & 10  & 3.0 & 1.6 & 38 & 52.1 &  1.6  \\
\hline
\sidehead{Fornax}
\hline
King  & 0.0 & 10  & 57.9 & 10.7    & 66 & 172.5 & 9.3 \\
$f(E, L_{z'})$ & 0.15 & 10 & 13.1 & 0.7    & 66 & 39.9 & 2.3  \\
$f(E, L, L_{z'})$ & 0.30 & 10 & 12.9  & 0.6    & 66 & 34.9 & 2.7 \\
\enddata
\end{deluxetable}


\begin{thebibliography}{99}

\bibitem[\protect\citeauthoryear{Aaronson}{1983}]{Aar83}Aaronson, M., 1983, \apj, 266, L11
\bibitem[\protect\citeauthoryear{Belokurov et al.}{2007}]{Bel07}Belokurov, V., 2007, \apj, 654, 897 
\bibitem[\protect\citeauthoryear{Binney \& Tremaine}{1987}]{Bin87}Binney, J., \& Tremaine S., 1987, Galactic Dynamics, Princeton University Press
\bibitem[\protect\citeauthoryear{Cretton et al.}{1999}]{Cre99}Cretton, N., De Zeeuw, P. T., van der Marel, R. P., \& Rix, H.-W., 1999, \apj, 124, 383
\bibitem[\protect\citeauthoryear{G{\'{o}}mez-Flechoso, Fux \& Martinet}{1999}]{Go99}Gomez-Flechoso, M. {\'{A}}., Fux, R. \& Martinet, L., 1999, \aa, 347, 77
\bibitem[\protect\citeauthoryear{G{\'{o}}mez-Flechoso \& Mart\'{i}nez-Delgado}{2003}]{Go03}G{\'{o}}mez-Flechoso, M. {\'{A}}, \& Mart\'{i}nez-Delgado, D., 2003, \apj, 586, L123
\bibitem[\protect\citeauthoryear{Ibata, Gilmore \& Irwin}{1994}]{Iba94}Ibata, R. A., Gilmore, G., \& Irwin, M. J., 1994, \nat, 370, 194
\bibitem[\protect\citeauthoryear{Ibata et al.}{2001}]{Iba01}Ibata, R., Irwin, M., Lewis, G. F., \& Stolte, A., 2001, \apj, 547, L133
\bibitem[\protect\citeauthoryear{Irwin \& Hatzidimitriou}{1995}]{IH95}Irwin, M., \& Hatzidimitriou, D., 1995, \mnras, 277, 1354
\bibitem[\protect\citeauthoryear{King}{1966}]{King66} King, I. R., 1966, \aj, 71, 1
\bibitem[\protect\citeauthoryear{Klessen et al.}{2003}]{Kle03}Klessen, R. S., Grebel, E. K., \& Harbeck, D., 2003, \apj, 589, 798
\bibitem[\protect\citeauthoryear{Kleyna et al.}{2002}]{Kle02}Kleyna, J., Wilkinson, M., I., Evans, N. W., Gilmore, G., \& Frayn, C., 2002, \mnras, 330, 792 
\bibitem[\protect\citeauthoryear{Kleyna et al.}{2004}]{Kleyna04}Kleyna, J. T., Wilkinson, M. I., Evans N. W., \& Gilmore, G., 2004, \mnras, 354, L66
\bibitem[\protect\citeauthoryear{Kleyna et al.}{2005}]{Kle05}Kleyna, J. T., Wilkinson, M. I., Evans N. W., \& Gilmore, G., 2005, \apj, 630, L141
\bibitem[\protect\citeauthoryear{Klypin et al.}{1999}]{Kly99}Klypin, A., Kravtsov, A. V., \& Valenzuela, O., 1999, \apj, 522, 82
\bibitem[\protect\citeauthoryear{Kroupa}{1997}]{Kro97}Kroupa, P., 1997, \na, 2, 139
\bibitem[\protect\citeauthoryear{{\L}okas}{2002}]{Lok02}{\L}okas, E. L., 2002, \mnras, 333, 697
\bibitem[\protect\citeauthoryear{{\L}okas et al.}{2005}]{Lok05}{\L}okas, E. L., Mamon, G. A., \& Prada, F., 2005, \mnras, 363, 918 
\bibitem[\protect\citeauthoryear{Mart\'{i}nez-Delgado et al.}{2001}]{Mar01}Mart\'{i}nez-Delgado, D., Alonso-Garc\'{i}a, J., Aparicio, A., \& G\'{o}mez-Flechoso, M. A., 2001, \apj, 549, L63
\bibitem[\protect\citeauthoryear{Mashchenko, Sills \& Couchman}{2004}]{Mas04}Mashchenko, S., Sills, A., \& Couchman, H. M. P., 2006, \apj, 640, 252
\bibitem[\protect\citeauthoryear{Mateo}{1998}]{Mat98}Mateo, M., 1998, \araa, 36, 435
\bibitem[\protect\citeauthoryear{Milgrom}{1983}]{Mil83}Milgrom, M., 1983, \apj, 270, 365
\bibitem[\protect\citeauthoryear{Mu\~{n}oz et al.}{2005}]{Mun05}Mu\~{n}oz, R., Frinchaboy, P. M., Majewski, S. R., Kuhn, J. R., Chou, M.-Y., Palma, C., Sohn, S. T., Patterson, R. J., \& Siegel, M. H., 2005, \apj, 631, L137 
\bibitem[\protect\citeauthoryear{Navarro, Frenk \& White}{1997}]{Nav97}Navarro, J. F., Frenk, C. S., \& White, S. D. M., 1997, \apj, 490, 493
\bibitem[\protect\citeauthoryear{Odenkirchen et al.}{2001}]{Ode01}Odenkirchen, M., et al., 2001, \aj, 122, 2538 
\bibitem[\protect\citeauthoryear{Olszewski, Pryor \& Armandroff}{1996}]{Ols96}Olszewski, E. W., Pryor, C., \& Armandroff, T. E., 1996, \aj, 111, 2
\bibitem[\protect\citeauthoryear{Palma et al.}{2003}]{Pal03}Palma, C., Majewski, S. R., Siegel, M. H., Patterson, R. J., Ostheimer, J. C., \& Link, R., 2003, \apj, 123, 1352
\bibitem[\protect\citeauthoryear{Piatek et al.}{2002}]{Pia02}Piatek, S., Pryor, C., Armandroff, T. E., \& Olszewski, E. W., 2002, \apj, 123, 2511
\bibitem[\protect\citeauthoryear{Read et al.}{2006}]{Rea06}Read, J., I., Wilkinson, M. I., Evans, N. W., Gilmore, G., \& Kleyna, J., 2006, \mnras, 367, 387
\bibitem[\protect\citeauthoryear{Spergel et al.}{2006}]{Spe06}Spergel, D. et al., 2006, astro-ph/0603449 
\bibitem[\protect\citeauthoryear{Trager et al.}{1995}]{Tra95}Trager, S. C., King I. R., \& Djorgovski, S., 1995, \aj, 109, 1
\bibitem[\protect\citeauthoryear{Wang et al.}{2005}]{Wan05}Wang, X., Woodroofe, M., Walker, M. G., \& Mateo, M., 2005, \apj, 626, 145
\bibitem[\protect\citeauthoryear{Walker et al.}{2006}]{Wal06}Walker, M. G., Mateo, M., Olszewski, E. W., Bernstein, R. A., Wang, X., \& Woodroofe, M., 2006, \aj, 131, 2114
\bibitem[\protect\citeauthoryear{Wilkinson et al.}{2002}]{Wil02}Wilkinson, M. I., Kleyna, J., Evans, N. W., \& Gilmore G., 2002, \mnras, 330, 778
\bibitem[\protect\citeauthoryear{Wilkinson et al.}{2004}]{Wil04}Wilkinson, M., Kleyna, J. T., Evans, N. W., Gilmore, G. F., Irwin M. J., \& Grebel, E. K., 2004, \apj, 611, L21
\bibitem[\protect\citeauthoryear{Wu \& Tremaine}{2006}]{Wu06} Wu, X., \& Tremaine, S., 2006, \apj, 643, 210


\end{thebibliography}
\end{document}